

\input harvmac
\noblackbox

\def \eq#1 {\eqno {(#1)}}

\def \ra {\rightarrow}
\def\np {  Nucl. Phys. }
\def \pl { Phys. Lett. }

\def \prl { Phys. Rev. Lett. }
\def \pr  { Phys. Rev. }


\def\a{\alpha}
\def\b{\beta}

\def\g{\gamma}

\def\d{\delta}

\def\e{\epsilon}

\def\p{\phi}

\def\m{\mu}
\def\n{\nu}

\def\s{\sigma}

\def \sm {$\s$-model\ }

\def \bd {\bar \del}

\def \ha {{1\over 2}}

\def \ov {\over}

\def \p {\phi}


\def \sms {$\s$-models\ }

\def \bd {\bar \del}

\def \ra {\rightarrow}

\def \a {\alpha}
\def \b {\beta}

\def \ln {{\rm \ ln \  }}

\def \p {\phi}

\def \m {\mu }
\def \n {\nu}

\def\g {\gamma}

\def \d {\delta}

\def \s {\sigma}

\def \fourth {{\textstyle{1\over 4}}}

\def \e#1 {{{\rm e}^{#1}}}

\def \eq#1 {\eqno {(#1)}}
\def \sm {$\s$-model\ }

\def \bd  {{ \bar \del }}

\def \bd  { \bar \del }


\def \p {\phi}

\def \s {\sigma}

\def \d {\delta}

\def \m {\mu}

\def \g {\gamma}
\def \n {\nu}

\def \fourth {{1\over 4}}

\def \e#1 {{{\rm e}^{#1}}}

\def \m {\mu}

\def \ra {\rightarrow}

\def \eq#1 {\eqno{(#1)}}
\def \e {\rm e}
\def \ra {\rightarrow }
\def \e#1 {{\rm e}^{#1}}

\def \ln { {\rm ln } }

\def \p {\phi}

\def  \g {\gamma}

\def\({\left (}
\def\){\right )}
\def\[{\left [}
\def\]{\right ]}

\def\np {  Nucl. Phys. }
\def \pl { Phys. Lett. }

\def \prl { Phys. Rev. Lett. }
\def \pr  { Phys. Rev. }

\def \jmp { Journ. Math. Phys. }

\baselineskip8pt
\Title{\vbox
{\baselineskip 6pt{\hbox{  }}{\hbox
{Imperial/TP/93-94/51 }}{\hbox {UCSBTH-94-24}}{\hbox{hep-th/9408040}} } }
{\vbox{\centerline {  Extremal black holes as exact  string solutions}
}}
\vskip 37 true pt
\centerline  { {Gary T. Horowitz\footnote {$^*$} {e-mail address:
gary@cosmic.physics.ucsb.edu} }}

 \smallskip \smallskip
\centerline {\it Physics
Department }
\smallskip

\centerline{\it  University of California, Santa Barbara, CA 93106, USA}
\medskip
\centerline {and}
\medskip
\centerline{   A.A. Tseytlin\footnote{$^{\star}$}{\baselineskip8pt
e-mail address: tseytlin@ic.ac.uk}\footnote{$^{\dagger}$}{\baselineskip5pt
On leave  from Lebedev  Physics
Institute, Moscow, Russia.} }

\smallskip\smallskip
\centerline {\it  Theoretical Physics Group, Blackett Laboratory}

\centerline {\it  Imperial College,  London SW7 2BZ, U.K. }
\bigskip
\centerline {\bf Abstract}
\medskip
\baselineskip12pt
\noindent
We show that
the leading order solution describing an extremal
electrically charged black hole in string theory is, in fact, an exact
solution to all orders in $\a'$
when interpreted in  a  Kaluza-Klein fashion.
 This follows from the observation  that it
can be obtained via dimensional
reduction from a five dimensional background which
 is proved to be an exact string solution.

\Date {August  1994}

\noblackbox
\baselineskip 20pt plus 2pt minus 2pt

\vfill\eject

\def \lr { \lref}

\gdef \jnl#1, #2, #3, 1#4#5#6{ {\sl #1~}{\bf #2} (1#4#5#6) #3}

\lr \horts { G. Horowitz and A. Tseytlin, ``On exact solutions and
singularities in string theory", preprint  Imperial/TP/93-94/38;
hep-th/9406067.}

\lr \hortsy{G. Horowitz and A. Tseytlin, ``New exact solutions in string
theory", in preparation.}
\lr \susy  {R. Kallosh, A. Linde, T. Ort\'in, A. Peet and A. Van Proeyen,  \jnl
\pr, D46, 5278, 1992.}

\lr \callan { C. Callan, R. Myers, and M. Perry, \jnl \np, B311, 673, 1988.}

\lr\rot{J.~Horne and G.~Horowitz, \jnl \pr, D46, 1340, 1992;
A. Sen, \jnl \prl, 69, 1006, 1992.}
\lr \ghrw{J. Gauntlett, J. Harvey, M. Robinson and D. Waldram,
\jnl \np, B411, 461, 1994.}
\lr \garf{D. Garfinkle, \jnl \pr, D46, 4286, 1992.}

\lr \gps {S.  Giddings, J. Polchinski and A. Strominger, \jnl  \pr,  D48,
 5784, 1993; W. Nelson, \jnl \pr, D49, 5302, 1994;
D. Lowe and A. Strominger, ``Exact Four-Dimensional Dyonic Black Holes and
Bertotti-Robinson Spacetimes
  in String Theory", UCSBTH-94-14, hep-th/9403186;
C. Johnson,  ``Exact models of extremal dyonic 4D black hole
solutions of heterotic string theory", IASSNS-HEP-94/20,
hep-th/9403192.}

\lr \hoho { J. Horne and G.  Horowitz, \jnl \np, B368, 444, 1992. }
\lr \horwel{G. Horowitz and D. Welch, \jnl \prl, 71, 328, 1993;
N. Kaloper,  \jnl \pr,  D48, 2598, 1993. }
\lr \host{ G. Horowitz and A. Steif,  \jnl \prl, 64, 260, 1990; \jnl \pr,
D42, 1950, 1990;  G. Horowitz, in: {\it
 Strings '90}, (eds. R Arnowitt et. al.)
 World Scientific (1991).}
\lr \busch {T.  Buscher, \jnl \pl, B194, 59, 1987; \jnl \pl,
 B201, 466, 1988.}
\lr \kallosh {E. Bergshoeff, I. Entrop, and R. Kallosh, ``Exact Duality in
String Effective Action", SU-ITP-93-37; hep-th/9401025.}

\lr \guv { R. G\"uven, \jnl \pl, B191, 275, 1987;
 D. Amati and C. Klim\v c\'\i k,
\jnl \pl, B219, 443, 1989;
 G. Horowitz and A. Steif,  \jnl \prl, 64, 260, 1990.}

\lr \duff {A. Chamseddine, M. Duff,  B. Nilsson, C. Pope and N. Warner,
 \jnl \pl,
B193, 444, 1987; D. Karabali, Q-Han Park and H. Schnitzer,
\jnl \pl, B205, 267, 1988; \jnl \np, B323, 572, 1989.}

\lr \gibb{A. Dabholkar, G. Gibbons, J. Harvey and F. Ruiz, \jnl \np, B340,
33, 1990.}

\lr \hhs{J. Horne, G. Horowitz and A. Steif, \jnl \prl, 68, 568, 1992.}

\lr \horstr{G. Horowitz and A. Strominger, \jnl \np, B360, 197, 1991.}

\lr \gib { G. Gibbons, \jnl \np, B207, 337, 1982. }

\lr \gim { G. Gibbons and K. Maeda,  \jnl \np, B298, 741, 1988. }
\lr\gar { D. Garfinkle, G. Horowitz and A. Strominger, \jnl \pr,  D43, 3140,
1991; {\bf D45} (1992) 3888(E). }
 \lr \witt{ E. Witten, \jnl \pr, D44, 314, 1991. }

\lr\kall { R. Kallosh, D. Kastor, T. Ort\'in   and T. Torma, ``Supersymmetry
and stationary solutions in dilaton-axion gravity",
SU-ITP-94-12, hep-th/9406059. }
\lr\joh { C. Johnson and R. Myers, ``Taub--NUT Dyons in Heterotic String
Theory",
IASSNS-HEP-94/50, McGill/94-28, hep-th/9406069. }

\lr\jons { C. Jonson and R.  Myers, ``Taub-NUT dyons in heterotic string
theory",
 IASSNS-HEP-94/50, hep-th/9406069. }

\lr \jon {C. Johnson,  ``Exact models of extremal dyonic 4D black hole
solutions of heterotic string theory", IASSNS-HEP-94/20,
hep-th/9403192.}

\lr \gurs{M. G\"urses, \jnl \pr, D46, 2522, 1992.}

\lr\bergsh { E. Bergshoeff, R. Kallosh and T. Ort\'in, \jnl \pr,  D47, 5444,
1993.}
\lr \wald {D. Waldram, \jnl \pr, D47, 2528, 1993.}

\lr \bergkal { E. Bergshoeff, R. Kallosh and T. Ort\'in, ``Black-hole-wave
duality in string theory",  SU-ITP-94-11, hep-th/9406009. }

\lr\sen{A. Sen, \jnl \np, B388, 457, 1992. }

\lr \huwi{ A. Sen, \jnl \pr, D32, 2102, 1985;
C. Hull and  E. Witten, \jnl \pl, B160, 398, 1985. }

\lr \mukhi{S. Mukhi, \jnl \pl, B162, 345, 1985.}

\lr\iwp{W. Israel and G. Wilson, \jnl \jmp, 13, 865, 1972;
Z. Perj\'es, \jnl \prl, 27, 1668, 1971.}

\lr\maha{J. Maharana and J. Schwarz, \jnl \np, B390, 3, 1993.}

\lr\nats{M. Natsuume, ``Higher order corrections to the GHS string black hole",
NSF-ITP-94-66, hep-th/9406079.}

\lr \nels {W. Nelson, \jnl \pr, D49, 5302, 1994.}

\lr \chs {C. Callan, J. Harvey and A. Strominger, \jnl \np,  B359,
 611,  1991.}

\lr \khga {R. Khuri, \jnl \np, B387, 315, 1992;  \jnl \pl, B294,
325, 1992; J. Gauntlett, J. Harvey and J. Liu, \jnl \np, B409, 363, 1993.}

\lr \dukh {M. Duff, R.  Khuri,  R. Minasyan and J. Rahmfeld,
\jnl \np, B418, 195, 1994. }

\lr \napp { M. McGuigan, C. Nappi and S. Yost, \jnl \np,   B375, 421,  1992. }
\lr \sfet{  K. Sfetsos and A. Tseytlin,  ``Four Dimensional Plane Wave String
Solutions with Coset CFT Description",
Imperial/TP/93-94/28, hep-th/9404063.      }
\lr \sus{ P. Howe and G. Papadopoulos, \jnl \np, B289, 264, 1987. }

\lr \sss {T. Banks and L. Dixon, \jnl \np, B307, 93, 1988.}
\lr \gaun{J. Gauntlett, talk presented at the conference
``Quantum Aspects of Black Holes",
Santa Barbara, June 1993.}

One of the goals of string theory for many years has been to find
exact black hole solutions in four dimensions.     The first nontrivial
string ($\a'$) corrections to the
Schwarzschild  solution have been investigated \callan,
and solutions to the leading
order equations describing charged black holes have been found
\refs{\gib,\gim,\gar}.
Exact black hole
solutions have been constructed in two \refs{\witt}   and three \horwel\
dimensions
using (gauged) WZW models based on the group $SL(2,R)$. In addition,
using the result that the extremal limit of a black five brane \horstr\
is an exact superstring solution \chs, one can (trivially)
dimensionally reduce to obtain an exact
five dimensional extreme black hole.
This black hole has a magnetic type of charge associated
with the antisymmetric tensor field.
It has recently been shown that one can also
dimensionally reduce the exact solution
in \chs\ down to four dimensions \khga\ to obtain an
extreme magnetically charged black hole
\dukh.\foot{These  exact five  and four  dimensional black hole solutions
are asymptotically
flat.  In addition, there are  coset CFT or gauged WZW
 constructions of just the `throat' regions  of some
 four dimensional extreme magnetically charged dilatonic black holes
\gps. The fact that the form of the throat solution is unchanged under string
corrections was noticed earlier \gurs.}

We shall
present an exact solution describing an
extremal  four dimensional $electrically$ charged black hole
in  (super)string theory. There are several important differences between
this and earlier work. First, the electrically charged solutions are
qualitatively
different  from their magnetic analogues. The extremal magnetically charged
solutions found in four and five dimensions are  products of a timelike line
and
a euclidean solution. So the lorentzian nature of the solution plays no role.
The extremal electrically charged solution, on the other hand, has a nontrivial
timelike direction. Second,  the previous four and five
dimensional extreme black holes were shown to be exact solutions only in the
superstring (or heterotic string) theory. Extended supersymmetry
(on the world-sheet  related to  that on space-time) played a key role in the
argument  \chs\ that there are no $\a'$-corrections to the leading-order
backgrounds. The solution we will discuss, like the previous  two
 and
three- dimensional examples, is exact   already  in the bosonic string theory.

The extremal electric black holes we shall consider are known to be
supersymmetric \susy\ and  can
be lifted to ten dimensions in such a way that they are dual to
 special plane fronted
waves \bergkal.  However, {\it a priori}  these properties by themselves are
not
sufficient to establish that extreme black holes
are  solutions to all orders in $\a'$.
 We will
see that another special property of the extremal electric  black holes
(related to a particular chiral coupling of the string to the background) can
be used to show
that they are exact solutions.

To satisfy the constraint on the central charge, every solution with
an asymptotically flat
four dimensional space-time  must have a number of internal dimensions. One
could  assume that the full solution is a simple product of
the four dimensional one with a small torus. However,  one could equally well
assume a tilted product in which off diagonal components of the higher
dimensional fields give rise to four dimensional gauge fields. This is
the approach we will adopt. We will, in fact, consider mainly the closed
bosonic
string theory which has no fundamental gauge fields in the higher
dimensional space.

It suffices to consider one nontrivial extra dimension (with the remaining ones
taken to be a trivial product).
We will show that
a particular five dimensional bosonic string background is an  exact solution,
and  that
its dimensional
reduction yields
\eqn\didi{ ds^2 = - F^2 (r) dt^2 + dr^2 + r^2 d\Omega \ , \ \ \  \ \ \ \ \
A_t = F(r)\ , } $$
 \p(r) = \p_0 + \ha \ln F (r)\ , \ \ \ \ \  \   F\inv (r)=    1 + {M\ov r}\ ,
  $$
where $d\Omega$ denotes the metric of a unit 2-sphere, $A_\m$ is the gauge
field potential, $\p$ is the dilaton  and $M$ is twice
the ADM mass.
This background is precisely the extremal limit of the known leading order
solution describing an electric dilaton black hole. In this sense one
can say that the leading order extremal black hole solution is exact.\foot{
Due to the well known ambiguity in the form of the higher order terms
in the string equations of motion, the phrase ``exact solution" will be
understood to mean ``exact in a particular renormalization scheme". Different
schemes
are equivalent in the sense that they  are related by local field redefinitions
which do not change the physical $S$-matrix. }

Let us first  describe a simple application of using dimensional
reduction to find exact extremal black holes in string theory.
Consider the five dimensional plane fronted wave
\eqn\plwv{ ds^2 = du dv + K(r) du^2 + dr^2 + r^2 d\Omega\ , \ \ \ \ \ \ \  K(r)
= 1+{M\ov r}\ .  }
This background (with constant dilaton and no additional fields) is
an exact solution to bosonic string theory \guv. To reduce to four
dimensions with $u$ as the internal coordinate, we rewrite \plwv\ in the form
\eqn\pwred{ ds^2 = K ( du + \ha {K\inv } dv)^2 -
\fourth K\inv {dv^2 } + dr^2 + r^2 d\Omega\ . }
Letting $t=v/2$, the four dimensional (string frame) metric, gauge field and
scalar
($G_{uu}= e^{-2\s}$)  are thus
\eqn\exttt{ ds^2= - K\inv (r) dt^2 +  dr^2 + r^2 d\Omega \ , \ \
\  A_t = K\inv (r)\ , \ \  \ \s =  -{\ha} \ln K (r) \ . }
This is just the extremal electrically charged Kaluza-Klein black hole
\refs{\gib,\dukh}.
So the extreme Kaluza-Klein black hole can be viewed as
an exact string solution.

We have recently shown \horts\ that a class of solutions to the leading
order bosonic
string equations are,  in fact,  exact. Included in this class was the five
dimensional fundamental string solution \gibb
\eqn\fs{ ds^2 = F(r) dudv + dr^2 + r^2 d\Omega\  ,  }
\eqn\bpf{B_{uv} = \ha F(r)\ ,\qquad  \p = \p_0 + \ha \ln F(r)\ ,
\qquad F(r) = \(1+{M\ov r}\)^{-1} \ ,   }
where $B_{\m\n}$ is the antisymmetric tensor.
If one writes $v= y+t ,\ u=y-t$ and reduces this solution
to four dimensions by assuming
$y$ to be  the internal coordinate, one again obtains the extreme Kaluza-Klein
black hole \exttt\ \gaun.
This is not surprising since the fundamental string solution
\fs\
is related to the plane fronted wave \plwv\ by a spacetime duality
transformation
\hhs,
and it turns out that  the Kaluza-Klein metric is  invariant  under this
duality.
The main effect of the duality transformation
is that the gauge field in four dimensions now
comes from the antisymmetric tensor $B_{\mu\nu}$ and not from the
off-diagonal components of the metric.

To obtain the extreme black hole  \didi\ in four dimensions, we must start with
the following generalization of the fundamental string solution
\eqn\gfs{ ds^2 = F(r) dudv + du^2+  dr^2 + r^2 d\Omega\ ,  }
with $B_{uv},\ \p$ and $F$ again given by \bpf. This  background
was found to be a solution of
the leading order string equations in \wald\ and was further discussed in
\ghrw. We will show that the arguments in \horts\
can in fact be extended to establish that \gfs\ is also an exact solution.
But first,
we describe some properties of this solution and
show that its dimensional reduction yields the extreme charged black hole.

Unlike the fundamental string \fs, the metric \gfs\
has a regular ergosphere. This can be seen by introducing
new coordinates $u= y-t$, $v=2t$ so that the metric becomes
\eqn\hgfs{ ds^2 = -\({r-M \ov r+M}\) dt^2 - {2M\ov r+M} dt dy + dy^2 +dr^2
      +r^2 d\Omega\ . }
It is now clear that the Killing vector $\partial /\partial  t$, which is a
unit time translation at infinity, becomes null at $r=M$. This surface is
an ergosphere and not an event horizon since it is timelike. One can still
travel from $r<M$ to $r>M$ provided one  moves in the $y$ direction.
Although the metric components remain finite at $r=0$, the metric becomes
degenerate there and the curvature diverges.

One can understand the
fact that \hgfs\ has an ergosphere while \fs\ does not as follows. It is
known  that the fundamental string is the extremal limit
of a two parameter family of charged black string solutions which can be
obtained by boosting the direct product of the
Schwarzschild background with a  line,
 and applying a duality
transformation \refs{\horstr,\hhs}. The result is
($S \equiv \sinh \a, \ C\equiv \cosh \a, $  $\  \a$ is the original boost
parameter)
\eqn\bksth{
 ds^2 = \(1+ {2mS^2\ov r}\)^{-1} \[-\(1-{2m\ov r}\) dt^2 + dy^2\]
      + \(1-{2m\ov r}\)^{-1} dr^2 + r^2 d\Omega \ , }
$$  B_{yt} = { C\ov S}  \(1 + {2mS^2\ov r}\)\inv
  \  ,\qquad e^{-2 \phi} = 1+{2m S^2\over r} \ .  $$
 The extremal limit
corresponds to  sending $m\ra 0, \ \a \ra \infty$ in such  a way that $ M\equiv
2me^{2\a}$
is held fixed. In this limit the horizon at $r=2m$ shrinks down to zero
size and becomes singular. The charged black string solution \bksth\ approaches
the fundamental string \fs. If we add linear momentum to \bksth\ by applying
a boost
$t=\hat t \cosh \b + \hat y \sinh \b, \ \ y=\hat t \sinh \b + \hat y \cosh \b$,
then we create an ergosphere outside  the horizon. The extremal limit
$m\ra 0, \ \a,\b \ra \infty$ with
$ M\equiv 2me^{2\a}=2me^{2\b}$ fixed, now yields the generalized fundamental
string \hgfs. So this solution can also be viewed as the extremal limit
of a charged black string, but now with non-zero  linear momentum.
The horizon still shrinks to zero size, but the ergosphere remains.

If $y$ is periodically identified with a small period, the generalized
fundamental string \hgfs\ will appear as four dimensional.
We can obtain the effective four
dimensional geometry by rewriting \hgfs\ in the form
\eqn\redfs{ ds^2 = \(dy - {M\ov r+M}dt\)^2 - {r^2 \ov (r+M)^2} dt^2 + dr^2
	 + r^2 d\Omega\ . }
Thus the four dimensional metric is
\eqn\fdson{ ds^2 = -{r^2 \ov (r+M)^2} dt^2 + dr^2 + r^2 d\Omega\ ,  }
i.e. is precisely that of the extreme electrically charged black hole \didi.
Note that the dilaton is the same as in five dimensions since the ``modulus"
field from the dimensional reduction is constant. The gauge field now
comes from both the off-diagonal components of the metric and the antisymmetric
tensor
which are equal (up to a gauge transformation).

We now turn to the demonstration that  the background \bpf,\gfs\ (or \hgfs) is
an exact string solution.
The bosonic string in a `massless' background is described  (in the conformal
gauge)
by the \sm
\eqn\mod{  I= {1\ov \pi \a' } \int d^2 z \ L\ , \ \ \ \ \ L=  (G_{\m\n} +
B_{\m\n})(X)\ \del X^\m \bd X^\n
+ \a'{\cal R}\p (X)\ ,  }
where $G_{\m\n}$ is the spacetime metric,  and
${\cal R}$ is related to the world-sheet metric $\g$  and its
scalar curvature by  $ {\cal R} \equiv \fourth \sqrt \g R^{(2)}$.
Consider a \sm of the form
\eqn\mof{  L_F= \ F(x) \ \del u \bd v + \del x^i \bd x^i
+ \a'{\cal R}\p (x) \  .  }
It was shown in \horts\ that this model is conformally invariant to all
orders (in a particular  scheme) if
\eqn\eqq{ \del^2 F\inv=0 \ , \ \ \ \ \   \p=  \p_0 + \ha  \ln F(x)\ .   }
The fundamental string solution corresponds to the case of $F^{-1} = 1 + M/r$.
The key feature which  enables one  to establish the conformal invariance to
all
orders is the chiral coupling of the $u$ and $v$ fields to the background,
which is obtained by setting $G_{uv} = B_{uv}$.
This results in the two chiral currents associated with the null translational
symmetries.
It turns out that a single chiral current associated with a null symmetry
is sufficient to establish the
conformal invariance  of similar backgrounds to all orders  provided
the one-loop conditions are satisfied. This means that a much larger class
of leading-order solutions can be shown to be exact. The most general
situation will be discussed elsewhere \hortsy. Here we consider the following
class of \sms which includes \bpf,\gfs\ as a special case
\eqn\mofk{ L_{FK}=F(x) \del u \bd v +  K(x)  \del u \bd u  +  \del x^i \bd x^i
   + \a'{\cal R}\p (x)\ .  }
To find the  exact  conditions of conformal invariance we  follow \horts\  by
introducing   the source terms
($z$ denotes the two world-sheet coordinates)
\eqn\sou{ L_{source} =  V(z)\del\bd u + U(z)\del\bd v  +  X(z)\del\bd x  \  , }
and performing  the path integral over $v$. The resulting $\d$-function
sets $\del u = F^{-1} \del U$ (up to a zero mode which we absorb in $U$).
We arrive at the following effective $x$-theory
\eqn\mofak{ L_{FK}'=\del x^i \bd x^i   -  F\inv(x) \del U \bd V }
$$ + \
  K (x) F\inv (x) \del U  \bd  \del^{-1}  [ F\inv (x) \del U ]
+ \a'{\cal R}(\p -\ha \ln F)  + \  X\del\bd x \   ,  $$
where the shift in $\p$ comes from the determinant and we make use
of a special scheme to keep the free kinetic term of $x^i$ unchanged (see
\horts).
The conditions of conformal invariance of the $O(\del U \bd V )$
term are the same as for the model with $K=0$   \eqq.
The conformal anomaly must be local, so that only the local part
of the non-local  $O(\del U \del U)$  term may contribute to it.
Since this non-local term  already contains two factors of $\del U$
it cannot produce  $\del X$-dependent counterterms. That means  we may
expand the functions $KF\inv\equiv\tilde K $ and $F\inv$ in it near a constant,
 $\ x^i(z)= x^i_0 + \eta^i(z) $
$$ \int d^2 z d^2z' [ \tilde K (x)  \del U](z)  {\bd^2  \Delta\inv (z, z') } [
F\inv (x) \del U ]  (z') $$ $$
=   \sum_{n,m=0}^{\infty} {1\ov n! m!}  \del_{i_1}...\del_{i_m}  \tilde K (x_0)
\del_{j_1}...\del_{j_n} F\inv (x_0) $$
 \eqn\xxx{\int d^2 z d^2 z' (\eta^{i_1} ... \eta^{i_m})(z) \del U (z) {\bd^2
 \Delta\inv (z, z') }  (\eta^{j_1} ... \eta^{j_n})(z') \del U (z') \ , }
where we defined $\Delta\inv$ by $\del\bd \Delta\inv = \delta^{(2)} (z,z')$.
Then the only contractions of the quantum fields $\eta^i$ that can produce
local
$O(\del U \bd U)$
 divergences  are the one-loop tadpoles on the left and right side of the
non-local
propagator $ \Delta\inv (z, z')$.
Any contraction between $\eta^n (z)$ and $\eta^m (z')$
gives  additional $ \Delta\inv (z, z')$-factors and thus contributes only
to the non-local part of the corresponding $2d$ effective action.

As a result, we find the following conformal invariance condition
\eqn\conn{  F\inv \del^2 \tilde  K   + \tilde K\del^2 F\inv =0 \ , }
or,  combined  with \eqq,
\eqn\eqqq{   \del^2 F\inv=0 \ , \ \ \  \del^2 (KF\inv) =0 \ , \ \ \   \p=  \p_0
+ \ha  \ln F(x)\ .   }
When $K=0$ these conditions obviously reduce to \eqq. When $F=1$ the \sm
\mofk\ describes the standard plane fronted wave, and \eqqq\ gives
 the usual $\del^2 K=0$ condition for this background.
The five dimensional solution \gfs\ which
yields the charged black hole corresponds to the simplest nontrivial
generalization \mofk\  of the fundamental string \mof\ with $K=1$.

This class of solutions is mapped into itself under the  duality transformation
with respect to a general spacelike translation in the $(u,v)$ plane. Setting
$v= \hat v + q u$ ($q=$const)
in \mofk\ and dualizing with respect to $u$ yields a \sm\
of exactly the same form with $F$, $K$, and $\p$ replaced by
\eqn\dudu{
\tilde F= {F\ov K + qF}\ , \ \ \  \tilde K= {1\ov K + qF}, \ \ \
   \tilde \p = \p - \ha \ln (K+qF)\  . }
In other words, the null translational symmetry and chiral coupling
are preserved under the duality. The conformal invariance
conditions \eqqq\ are of course invariant under this transformation.


The  above discussion  was in the context of the bosonic string theory.
A generalization to the case of the closed superstring theory is
straightforward.
One only has to repeat our arguments starting with the
 $(1,1)$ supersymmetric extension of the bosonic $\s$-model \mofk\
(and to note that the one-loop conformal invariance conditions are the same
as in the non-supersymmetric  case)
with the conclusion that the above bosonic backgrounds represent
also the superstring solutions.
 Moreover,
any closed superstring solution can  be embedded into  the heterotic
 string theory by
identifying the  generalized
spin connection with  a heterotic gauge field background
(i.e.  by   rewriting the $(1,1)$ supersymmetric \sm  in the
$(1,0)$-supersymmetric heterotic  form \huwi).

 Therefore, the exact   bosonic  solution \bpf,\gfs\
is also an  exact $D=5$ solution of the heterotic string theory.
Like the bosonic one,
it can also  be  given a  Kaluza-Klein interpretation  as a  four dimensional
extremal electric black hole background.
In this case the  four dimensional abelian  gauge field of the  electric
dilatonic
black hole  has a Kaluza-Klein origin, while the  role  of the
heterotic gauge field  background is to  combine with the  spin connection
contributions
to make the model $(1,1)$ supersymmetric.\foot{Note that  the  Yang-Mills
Tr$F^2\ $-term in the four dimensional  heterotic string effective action is
then treated as an $\a'$-correction (i.e.  on  an  equal footing with the
$R^2$-term)
while the abelian  $F^2$-term  originating from the $D=5$ scalar curvature term
 has the same order as the  $D=4$ Einstein $R$-term.  }

Given a $D=4$ leading-order bosonic  background, its  embedding
into the heterotic string theory is not unique.
The  embeddings of extremal $D=4$ dilatonic black holes
in which the $U(1)$ gauge field has Kaluza-Klein and not  heterotic
Yang-Mills origin
have extended ($N=2, \ D=4$) space-time supersymmetry \susy.
This suggests that the corresponding world-sheet theory should presumably  have
$(4,4)$ supersymmetry \sss\ and thus, as in the case  considered in \chs,  the
leading-order higher-dimensional superstring or heterotic string solutions
should remain exact (both in the  magnetic and electric cases).\foot{This
possibility was advocated to us by A. Strominger.}

At the same time, there   should  also exist  a related
solution of the  heterotic string theory formulated directly in $D=4$.
In  fact,  the charged dilatonic black hole  may be considered as   a
non-supersymmetric leading-order solution of the  $D=4$ heterotic string theory
with the charge
being that of the $U(1)$ subgroup of the  Yang-Mills  gauge group.
This solution must have an  extension to
higher orders in $\a'$  which, in general,  may not be the same
as the  above `Kaluza-Klein' solution.
Even though the leading-order terms in the
 compactified (from $D=5$ to $D=4$) bosonic
 string theory and $D=4$ heterotic string theory
with a $U(1)$ gauge field background
look the same, the $\a'$-corrections are  different, so
that  our bosonic  result  does not  automatically imply  that the extremal
electric black hole  considered as a  $D=4$  heterotic string solution is also
exact.
However,
given that  the bosonic string solution is just the leading-order one  when
considered in a special scheme,
it  is an interesting possibility that  a  similar special scheme
exists also in the  heterotic case.
If true, that would imply that extremal electrically charged
black holes remain exact
heterotic string solutions even in the case of the
non-supersymmetric embedding.\foot{This does not seem to be true for the
magnetically charged black hole, which
has  non-trivial $\a'$ corrections in the non-supersymmetric  case \nats. }


To summarize, we have found an exact five dimensional string solution
\bpf,\gfs\
which reduces to the extremal electrically charged black hole \didi\ in four
dimensions.
Our results  further support the  special nature
of extremal black holes. In particular, the exactness of the leading
order extremal electric black hole implies   that the  extreme
charge to mass ratio
should not be renormalised by $\a'$ corrections.
 Our method  does not
 apply to non-extremal black holes which are
 likely to have $\a'$ corrections in all renormalization schemes.
 (It is easy to show that this is the case for the Schwarzschild solution.)

One of the
motivations for having exact black hole solutions
is to better understand the behavior of strong gravitational fields
and the possible existence
of singularities
in string theory. Some preliminary observations are the following. In the
bosonic string theory the solution
appears as four dimensional with a null singularity at low energy. However,
as one approaches the singularity, one discovers that the solution is
fundamentally the product of a five dimensional solution and a torus. The
five dimensional solution has an ergosphere with a curvature singularity
inside. Whether this singularity adversely affects string propagation remains
to be seen.
One should note that the string coupling $e^\p$ goes to zero near the
singularity suggesting that quantum corrections will be suppressed.\foot{
This can be compared with the extreme Kaluza-Klein black hole \exttt\
which, as  we have seen,
is also an exact solution. This solution appears as four dimensional
with a timelike (naked) singularity at low energy. But near the singularity
one again finds that the solution is actually a product of a torus with
a five dimensional solution which has an ergosphere
and a
singularity inside. However, in this case the string coupling  remains
constant. }

As we noted earlier, the methods used here can be applied to a larger
class of backgrounds to show that many low energy solutions are, in
fact,  exact. These include the recently discovered dilatonic IWP metrics
\refs{\kall, \joh}. The details will be given elsewhere \hortsy.

\bigskip
We wish to thank  G. Gibbons,  R. Kallosh  and A. Strominger for useful and
stimulating
discussions. Part of this work was done during the  LMS Durham Symposium
 ``Quantum concepts in time and space"   and we would like to express our
gratitude to the organizers of this excellent meeting.
G.H. was supported in part by NSF Grant PHY-9008502 and by EPSRC
grant GR/J82041.
A.A.T. acknowledges the support of PPARC.

\vfill\eject

\listrefs
\end